\documentclass{article}

\usepackage{arxiv}

\usepackage[utf8]{inputenc} 
\usepackage[T1]{fontenc}    
\usepackage{hyperref}       
\usepackage{url}            
\usepackage{booktabs}       
\usepackage{amsfonts}       
\usepackage{nicefrac}

\usepackage{epsfig}
\usepackage{subcaption}
\usepackage{calc}
\usepackage{amssymb}
\usepackage{amstext}
\usepackage{amsmath}
\usepackage{amsthm}
\usepackage{multirow}
\usepackage{tabularx}
\usepackage{multicol}
\usepackage{enumitem}
\usepackage{pslatex}
\usepackage{graphicx}
\usepackage{textcomp}
\usepackage[flushleft]{threeparttable}
\usepackage{cellspace}
\usepackage{scalerel}
\usepackage{mathtools}
\usepackage{multirow}
\usepackage{mathtools}
\usepackage{hyperref}
\usepackage{array}
\usepackage{graphicx}
\usepackage{booktabs}
\usepackage{array,booktabs}
\usepackage{makecell}
\usepackage{amsmath}
\usepackage{float}
\usepackage{textcomp}

\usepackage{microtype}      
\usepackage{lipsum}
\usepackage{graphicx}
\graphicspath{ {./images/} }

\title{Evaluating Explainable AI for Deep Learning-based Network Intrusion Detection System Alert Classification}

\author{
 Rajesh Kalakoti \\
  Department of Software Science\\
    Tallinn University of Technology,\\
  Tallinn, Estonia \\
  \texttt{rajesh.kalakoti@taltech.ee} \\
   \And
 Risto Vaarandi \\
  Department of Software Science\\
    Tallinn University of Technology,\\
  Tallinn, Estonia \\
  \texttt{risto.vaarandi@taltech.ee} \\
  \And
Hayretdin Bahşi \\
  School of Informatics, Computing, and Cyber Systems\\
  Northern Arizona University\\
  United States\\
  \texttt{hayretdin.bahsi@taltech.ee} \\
    \And
Sven N\~omm \\
  Department of Software Science\\
    Tallinn University of Technology,\\
  Tallinn, Estonia \\
  \texttt{sven.nomm@taltech.ee} \\
}

\begin{document}
\maketitle
\begin{abstract}
A Network Intrusion Detection System (NIDS) monitors networks for cyber attacks and other unwanted activities. However, NIDS solutions often generate an overwhelming number of alerts daily, making it challenging for analysts to prioritize high-priority threats. While deep learning models promise to automate the prioritization of NIDS alerts, the lack of transparency in these models can undermine trust in their decision-making. This study highlights the critical need for explainable artificial intelligence (XAI) in NIDS alert classification to improve trust and interpretability. We employed a real-world NIDS alert dataset from 
Security Operations Center (SOC) of TalTech (Tallinn University Of Technology) 
in Estonia, developing a Long Short-Term Memory (LSTM) model to prioritize alerts. To explain the LSTM model's alert prioritization decisions, we implemented and compared four XAI methods: Local Interpretable Model-Agnostic Explanations (LIME), SHapley Additive exPlanations (SHAP), Integrated Gradients, and DeepLIFT. The quality of these XAI methods was assessed using a comprehensive framework that evaluated faithfulness, complexity, robustness, and reliability. Our results demonstrate that DeepLIFT consistently outperformed the other XAI methods, providing explanations with high faithfulness, low complexity, robust performance, and strong reliability. In collaboration with SOC analysts, we identified key features essential for effective alert classification. The strong alignment between these analyst-identified features and those obtained by the XAI methods validates their effectiveness and enhances the practical applicability of our approach. 
\end{abstract}



\section{\uppercase{Introduction}}\label{sec:introduction}

Many organizations use open-source (e.g., Suricata and Snort) or commercial (e.g., Cisco NGIPS) NIDS platforms to identify malicious network traffic~\cite{day2011performance}. Most widely used NIDS platforms use human-created signatures to identify malicious network traffic. However, this often results in many alerts, with only a tiny fraction deserving closer attention from security analysts~\cite{jyothsna2011review}. 

In a typical SOC operation, security analysts analyze the alerts based on their impact on the security of the organizational assets and categorize them as high or low priority. At this stage, analysts also identify the false positives that are benign system activities but are flagged as alerts by NIDS. Security analysts find it challenging to identify high-priority alerts~\cite{jyothsna2011review}. Machine learning (ML) Deep Learning (DL) methods constitute a significant solution to automatize these prioritization tasks and, thus, reduce SOC workloads, especially in the lower-tier levels of security monitoring and incident handling processes in the related literature, with approaches divided into supervised, unsupervised, and semi-automated methods~\cite{vaarandi2021stream,vaarandi2022build,kalakoti2022depth}. However, the explainability or interpretability of ML models arises as a significant concern in alert prioritization despite their significant contribution. 

Explainable Artificial Intelligence (XAI or Explainable AI) is necessary for experts to verify alert classifications and for industries to comply with regulations~\cite{goodman2017european}. In cybersecurity, it's vital to explain flagged network activities as potential threats. XAI helps meet compliance standards and improve systems by clarifying NIDS alert classifications and identifying crucial features for data collection.  In the event of a security breach, XAI offers valuable insights for forensic analysis, helping to understand why specific alerts were or were not triggered, which is crucial in reconstructing the timeline and nature of an attack~\cite{alam2024xai}.  NIDS usually struggles with high false positive rates. XAI can enable security analysts to understand why particular benign activities are mistakenly flagged as threats, enabling more transparent system tuning and reducing false alarms~\cite{moustafa2023explainable}. 

Explainable AI (XAI) methods address the model opacity problem through various global and local explanation methods \cite{rawal2021recent}. Several studies have studied explainable AI methods in intrusion detection~\cite{alam2024xai,szczepanski2020achieving,senevirathna2024deceiving,moustafa2023explainable}. However, it is crucial to note that these studies did not comprehensively evaluate Explainable AI methods under various intrusion datasets and miscellaneous sets of Black box nature of AI  models. This lack of comprehensive evaluation significantly affects the generality of such methods, highlighting the urgent need for further research in this area. Although XAI-based IDS  tools are expected to be an integral part of network security to help security analysts in 
SOCs
to enhance the efficiency and precise in network defence and threat mitigation, a key challenge of deploying XAI-Based model into network intrusion detection is assessing such tools, testing their quality, and evaluating the relevant security metrics. These challenges undermine the trust in using the XAI-IDS model for real-world deployment in network IDS systems. 

In this paper, we propose a Long Short-Term Memory (LSTM) model for NIDS alert prioritization to improve transparency and Reliability. This study evaluates various XAI methods to bridge the gap between the high accuracy of complex ML models and the need for transparent, explainable decision-making in the cybersecurity problem domain. Objectives of the study include creating an explainable LSTM model for NIDS alert classification, comparing four advanced XAI methods, evaluating their performance using comprehensive metrics, and validating XAI-generated explanations based on four criteria: Faithfulness, Complexity, Robustness, and Reliability. 

Faithfulness estimates how accurately the explanation reflects the model’s behaviour, assuring that the local explanation represents the model’s decision-making process. Robustness evaluates the stability of explanations under small input perturbations, which is vital for building faith in local explanations. Complexity assesses the simplicity of the explanations, as more detailed explanations are generally more interpretable and valuable for human understanding. Reliability guarantees that the explanations are consistent with established knowledge, such as the features identified by SOC analysts in this case. 

We propose that explainable AI methods can provide explanations for the decision-making processes of the LSTM  model, prioritizing NIDS 
alerts and ultimately boosting the trust and usefulness of these systems. This research particularly examined a real-world dataset of NIDS alerts using LSTM, interpreting the output decisions made by these models and evaluating them through both quantitative and qualitative (expert) evaluations. This study emphasizes artificial intelligence (XAI) in high-risk threat detection settings. Our research offers a perspective to the existing literature as the aspect of interpretability has not been explored in relation to the significance of NIDS alerts.  This research suggests that a well-designed benchmarking study can identify high-performance detection models that provide high-quality explanations. Therefore, security experts may not need to sacrifice detection performance over a model for explainability in the addressed ML studies.

Our paper is structured as follows: Section~\ref{sec:related_work} reviews related work on NIDS and XAI in NIDS, Section~\ref{sec:methodology} outlines our methodology, Section~\ref{sec:results_discussion} presents our results and discussions, and Section~\ref{sec:conclusions} provides our conclusions.

\section{Related Work}\label{sec:related_work}
ML and DL have advanced the analysis of NIDS alerts.
This section reviews key contributions in NIDS alert processing, focusing on classification, clustering, and explainable AI methods. It delves into studies addressing challenges such as alert prioritization, false positive reduction, and interpretable models in cybersecurity.

~\cite{kidmose2020featureless} proposed a three-phase method for NIDS alert classification~\cite{kidmose2020featureless}. They used an LSTM and latent semantic analysis to convert textual alerts into vectors, clustered the vectors using the DBSCAN algorithm, and classified incoming alerts based on their similarity to the core points of the clusters. \cite{van2022deepcase} developed a semi-automated method for classifying NIDS alerts and other security events,which involved detecting and analyzing event sequences using deep learning models, clustering with the DBSCAN algorithm, and human analysts labeling the resulting clusters~\cite{van2022deepcase}. Labeled database was then used for semi-automated classification of additional event sequences, with human analysts manually reviewing unclustered events.

In a paper\cite{mane2021explaining}, the authors utilized SHAP, LIME, Contrastive Explanations Method (CEM), ProtoDash, and Boolean Decision Rules via Column Generation (BRCG) over the NSL-KDD dataset~\cite{tavallaee2009detailed} for intrusion detection system (IDS). They demonstrated the factors that influence the prediction of cyber-attacks.

~\cite{ban2023breaking} proposed a method using an IWSVM-based classifier to detect critical NIDS alerts. The classifier assigned higher weights to repeated data points and the minority class of critical alerts. A clustering algorithm grouped alerts representing the same incident based on attributes such as IP addresses, service ports, and alert occurrence time.~\cite{shin2019platform} developed an organizational platform using machine learning to analyze NIDS alert data with support for binary SVM and one-class SVM methods~\cite{shin2019platform}. In a paper~\cite{feng2017user}, authors described another organizational implementation for processing NIDS alerts and other security events to identify at-risk users.~\cite{wang2019identifying} used a graph-based method to eliminate false alerts and applied GBDT algorithms for alert classification.~\cite{ban2021combat} used a large NIDS dataset to evaluate seven supervised machine learning methods~\cite{ban2021combat}. They found that Weighted SVM, SVM, and AB (Adaboost) produced the best results, while two isolation forest-based unsupervised algorithms provided lower precision than the evaluated supervised algorithms. 

It is important to note that a large body of research is devoted to replacing NIDS with ML-based systems \cite{tsai2009intrusion}. However, organizations use signature-based NIDSs due to the wide availability of this technology and complex SOC processes evolving around these systems. Thus, prioritizing NIDS alerts is a significant real-world challenge in SOCs. Various research studies have addressed the explainability of ML-based NIDS systems. However, to our knowledge, the explainability of the ML models developed for NIDS alert prioritization has not been studied in the literature.

\cite{szczepanski2020achieving} introduced the hybrid Oracle Explainer IDS, which combines artificial neural networks and decision trees to achieve high accuracy and provide human-understandable explanations for its decisions~\cite{szczepanski2020achieving}. In a paper~\cite{senevirathna2024deceiving}, authors have developed an Oracle-based Explainer module that uses the closest cluster to generate an explanation for the decision. A study explores how explanations in the context of 5G security can be targeted and weakened using scaffolding techniques. The authors suggest a framework for carrying out the scaffolding attack within a security setting, which involves selecting features and training models by combining explainable AI methods. \cite{zolanvari2021trust}\cite{zolanvari2021trust} introduced a model-agnostic XAI framework called TRUST for numerical applications. It uses factor analysis to transform input features, mutual information to rank features, and a multimodal Gaussian distribution to generate new samples for each class label.  

Some other studies have explored explainable AI methods in intrusion detection~\cite{alam2024xai,szczepanski2020achieving,kumar2024evaluating,kalakoti2024improving,kalakoti2024enhancing,kalakoti2024fedxai,kalakoti2023improving}. In contrast to studies on machine learning-based Network Intrusion Detection Systems (NIDSs), our research emphasizes the significance of making NIDS alerts understandable through model transparency. Our approach incorporates eXplainable AI (XAI) techniques to evaluate their effectiveness in clarifying NIDS alert classifications. We worked with a real world NIDS dataset from an environment making our findings more relevant than those based on old data sets. Our evaluation criteria cover aspects such as the reliability, faithfulness, robustness and complexity of explanations assessing explainability within this domain.  By engaging Security Operations Center (SOC) analysts in verifying our XAI findings we bridge the gap, between machine learning models and human knowledge. This progress enhances XAI in the field of cybersecurity, offering perspectives for developing transparent and reliable NIDS alert critical prioritization systems.

\section{Methodology}\label{sec:methodology}
\subsection{Dataset}
Our study makes use of a NIDS alert dataset taken from a Suricata NIDS system deployed at the Security Operations Center (SOC) of Tallinn University of Technology (Taltech). The dataset was gathered using the Customized Stream Clustering Algorithm for Suricata (CSCAS) to analyze alerts from Suricata NIDS at TalTechs SOC. Data was collected over a span of 60 days, from January to March 2022 during which Suricata generated alerts, for network activity involving 45,339 hosts and 4401 TalTech hosts. 

Throughout the data collection phase CSCAS operated with settings; SessionLength = 300 seconds (5 minutes) SessionTimeout = 60 seconds (1 minute) ClusterTimeout = 604,800 seconds (1 week) CandTimeout = 36,000 seconds (10 hours) MaxCandAge = 864,000 seconds (10 days) and \(\alpha = 0.01\). These configurations have been employed for CSCAS in an environment since 2021 and were determined to be optimal as outlined in~\cite{vaarandi2021stream}. NIDS Alerts are classified as either "important" or "irrelevant." Data points of network traffic were generated by a customized version of SCAS, a stream clustering algorithm, and have labels indicating whether they are regarded as inliers or outliers by SCAS. Data points are labeled by humans to indicate if they represent important or irrelevant alert groups. Important alerts are prioritized in the SOC security monitoring processes. Irrelevant alerts include low-priority threats (e.g., frequent scanning for old vulnerabilities) or false positives (e.g., alerts related to attempts to resolve botnet C\&C server DNS names not originating from infected computers but from specific security applications). The description of the dataset~\cite{vaarandi2024dataset} is given below:
\begin{itemize}[leftmargin=2em, topsep=-0.5pt, partopsep=0pt, parsep=0pt, itemsep=0pt]
    \item Timestamp -- alert group reporting time
    \item SignatureText -- human readable alert text
    \item SignatureID -- numerical signature ID
    \item SignatureMatchesPerDay -- Average matches per day by the triggering signature (set to 0 if first match was less than 24 hours ago).
    \item AlertCount -- the number of alerts in the current alert group
    \item Proto -- numerical protocol ID (e.g., 6 denotes TCP and 17 UDP)
    \item ExtIP -- anonymized IP address of the external host (extipN, where N is a number that identifies the given IP address)
    \item ExtPort -- port at the external host, set to -1 if alerts involve multiple external ports
    \item IntIP -- Anonymized IP address of the internal host (intipN), set to -1 if alerts involved multiple internal IP addresses.
    \item IntPort -- port at the internal host, set to -1 if alerts involve multiple internal ports.
    \item Similarity -- The overall similarity of this alert group to others in the same cluster or, if it's an outlier, to other outlier alert groups. The value ranges from 0 to 1, with higher values indicating a high degree of similarity.
    \item SCAS -- The label assigned by the customized version of SCAS. Here, 0 denotes an inlier and 1 denotes an outlier.
    \item AttrSimilarity -- similarity for the network IDS alert attribute Attr (there are 34 attributes in total). Set to -1 if the attribute Attr is not set for the given signature, otherwise ranges from 0 to 1. The field indicates how often the attribute value has been observed in other alert groups from the same cluster (or in other outlier alert groups if the current alert group is an outlier).
\end{itemize}

We collaborated with Security Operations Center (SOC) analysts from TalTech, Estonia to estimate the reliability of the post-hoc explanations generated for the decisions of the black-box model, which is the DL model induced for alert classification in this work. A detailed description of TalTech SOC can be found in~\cite{vaarandi2022build}. Leveraging their expertise in managing Network Intrusion Detection System (NIDS) alerts, the SOC team at TalTech identified the five features for determining alert significance as outlined in Table \ref{tab:soc_features}. These features act as benchmarking reference features in our research to evaluate how well our XAI algorithms perform.

\begin{table}[ht]
\centering
\caption{Key Features Identified by Taltech SOC Analyst for Determining NIDS Alert Significance}
    \begin{tabular}{|l|}
    \hline
    SignatureMatchesPerDay \\ \hline
    Similarity             \\ \hline
    SCAS                   \\ \hline
    SignatureID            \\ \hline
    SignatureIDSimilarity  \\ \hline
    \end{tabular}
    \label{tab:soc_features}
\end{table}

For our work, the dataset excluded 'SignatureText' and 'Timestamp' features as external IP addresses ("ExtIP" feature) and internal IP addresses ("IntIP" feature) prior, to model training. 
\subsection{Long Short-Term Memory for NIDS alerts}
\begin{figure}[ht]
    \centering
    \setkeys{Gin}{width=\linewidth}
    \includegraphics{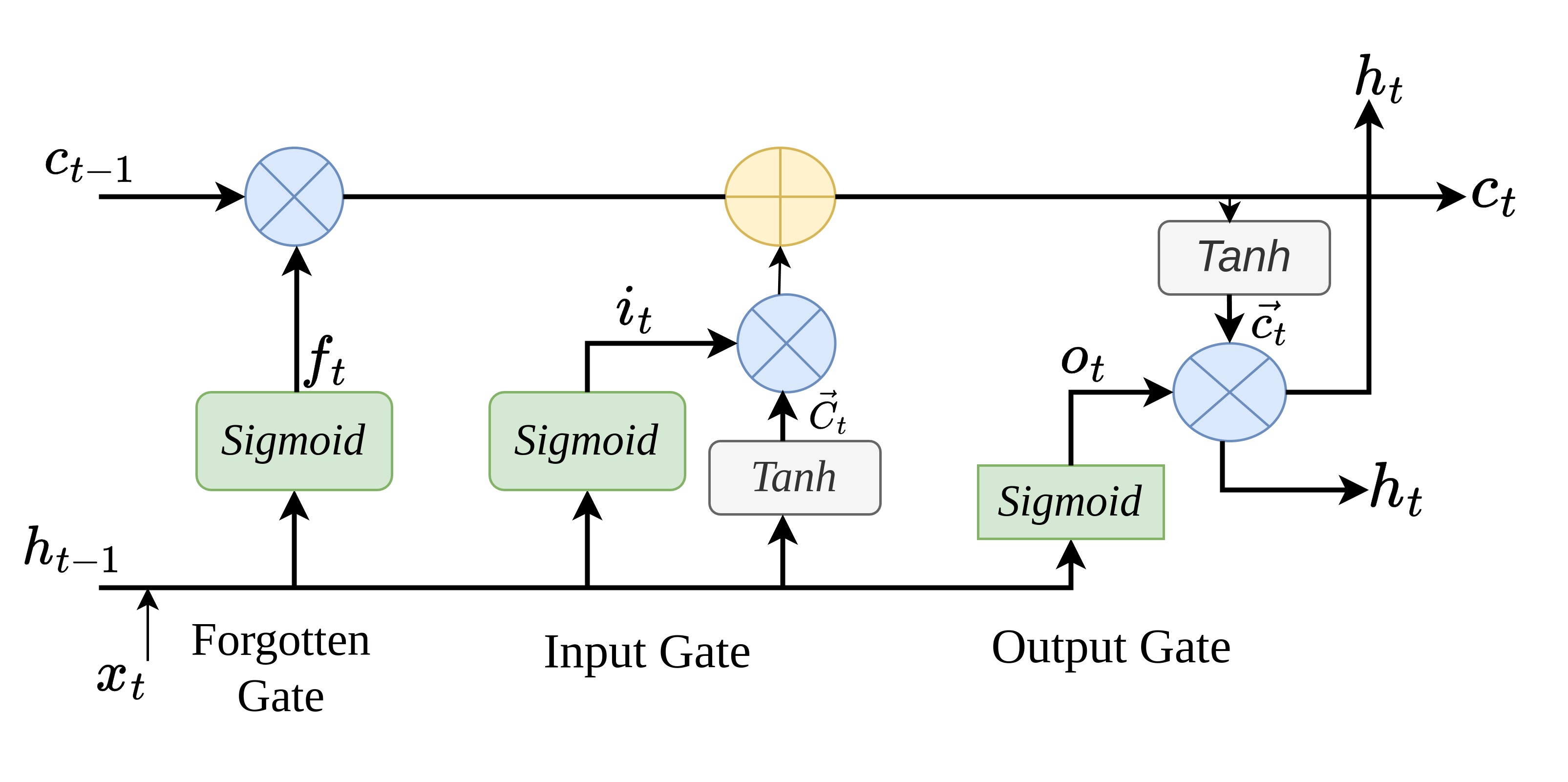}
    \caption{Hidden Layer Architecture of LSTM Network}
    \label{fig:lstm_model}
\end{figure}

In this study, we proposed long-term memory (LSTM) to classify whether a given NIDS alert group needs immediate attention (Important class label) or can be assessed as less critical (Irrelevant class label). LSTM is a neural network designed to address the long-term dependence problem in traditional recurrent neural networks. It introduces forget,  input, and output gates to control the flow of information and maintain long-term memory. Figure~\ref{fig:lstm_model} shows structure of the hidden layer of the LSTM network. The forget gate adapts to the context, discarding unnecessary information. It uses a sigmoid function to produce a value between 0 and 1, then multiplied by the previous cell state. A value of 0 means complete forgetting, while 1 means fully retained.

\begin{equation}
    f_t = \sigma (W_f \cdot [h_{t-1}, x_t] + b_f) 
\end{equation}

The input gate enhances the necessary information for the new cell state, and its output is a sigmoid function with a range of 0 to 1, which is multiplied by the current cell state.

\begin{equation}
    i_t = \sigma (W_i \cdot [h_{t-1}, x_t] + b_i) 
\end{equation}

\begin{equation}
    \tilde{C}_t = \tanh (W_c \cdot [h_{t-1}, x_t] + b_c) 
\end{equation}

Then the old and new state information can be combined to construct the final new cell state.

\begin{equation}
    C_t = f_t \times C_{t-1} + i_t \times \tilde{C}_t
\end{equation}

The output is determined by the output gate, which uses a sigmoid function to select information to be output along with the final cell state and the Tanh function. 
\begin{equation}
    O_t = \sigma (W_o \cdot [h_{t-1}, x_t] + b_o) 
\end{equation}
\begin{equation}
    h_t = O_t \times \tanh (C_t) 
\end{equation}

For training LSTM model, We selected 10,000 data points for each class label ('irrelevant' and 'important'), resulting in a total of 20,000 samples. The The dataset was divided into training and testing sets at an 80 20-split ratio. 
We applied the data normalization technique to the dataset to convert the values to a standard scale. We used Min-Max normalization, one of several available techniques, to transform and normalize the input features to scale them within a range of 0 to 1, as shown in Equation~\ref{eq:norm}.

\begin{equation}\label{eq:norm}
    x' = \frac{x - x_{\text{min}}}{x_{\text{max}} - x_{\text{min}}}
\end{equation}
where \(x_{\text{min}}\) is the smallest value of the feature, \(x_{\text{max}}\) is the largest value of the feature, and \(x\) is the actual value of the feature. The normalized feature, \(x'\), ranges between 0 and 1.

We used RandomSearch hyperparameter tuning with Ray Tune library\footnote{https://docs.ray.io/en/latest/tune/index.html} to train LSTM model.  We evaluated the performance of LSTM model for NIDS alerts classification using a confusion matrix. In NIDS alerts classification, True Positives (TP) are the number of important alerts correctly classified as important, True Negatives (TN) are the number of irrelevant alerts correctly classified as irrelevant, False Positives (FP) are the number of irrelevant alerts incorrectly classified as important. False Negatives (FN) are the number of important alerts incorrectly classified as irrelevant. we used the following evaluation metrics

\begin{equation}
    \text{Accuracy} = \frac{TP + TN}{TP + TN + FP + FN}
\end{equation}

\begin{equation}
    \text{Precision} = \frac{TP}{TP + FP}
\end{equation}

\begin{equation}
    \text{Recall} = \frac{TP}{TP + FN}
\end{equation}

\begin{equation}
    \text{F1-Score} = 2 \times \frac{\text{Precision} \times \text{Recall}}{\text{Precision} + \text{Recall}}
\end{equation} 
 
We used softmax activation function at the output layer to predict class labels, which provides prediction probabilities for each class and enables us to understand the model's confidence and the probability distribution. It's also crucial to evaluate XAI techniques based on metrics like faithfulness, monotonicity and max sensitivity as discussed in section \ref{sec:evaluate_xai}.

\subsection{Explainable AI Methods}
When explaining the model using Explainable AI, there are two approaches: model agnostic and model specific. Explainable AI methods are also categorized into two types explanations. Local explanations interpret individual predictions and global explanations that offer an overview of the model's behaviour. Our goal is to enhance the explainability of NIDS alerts detected by LSTM model. We have utilized four popular XAI feature attribution methods. Will provide a brief overview of each one. The following outlines the four methods (LIME, SHAP, Integrated Gradients (IG) and DeepLIFT) all designed to clarify instances and shed light on how the model makes decisions, for specific predictions. Let $x \in \mathbb{R}^d$ be the input, where $d$ is the feature set dimensionality. The black box model $\mathcal{M}$ maps input to output $\mathcal{M}(x) \in \mathcal{Y}$. Dataset $\mathcal{D} = {(x^i, y^i)}$ contains all input-output pairs. The explanation mapping $\mathbf{g}$ for predictor $\mathcal{M}$ and point $x$ returns importance scores $\mathbf{g}(\mathcal{M}, x) = \phi_x \in \mathbb{R}^d$ for all features. Let $D: \mathbb{R}^d \times \mathbb{R}^d \mapsto \mathbb{R}{\geq 0}$ be a metric in the explanation space and $S: \mathbb{R}^d \times \mathbb{R}^d \mapsto \mathbb{R}{\geq 0}$ a metric in the input space. The evaluation criterion $\mu$ maps predictor $\mathcal{M}$, explainer $\mathbf{g}$, and point $x$ to a scalar.

\subsubsection{SHAP}
SHAP~\cite{lundberg2017unified} uses Shapley values from game theory to attribute the importance of each feature to a model's prediction, providing a unified measure of feature importance.  SHAP based on Shapley values, is defined as: ${\bf g} ({\bf \mathcal{M}}, x) = \phi_0 + \sum_{j=1}^M \phi_j$
where $\phi_j$ is the feature attribution of feature $j$. SHAP's DeepExplainer was used in this study.  

\subsubsection{LIME} 
LIME~\cite{ribeiro2016should} (Local Interpretable Model-agnostic Explanations) constructs a locally interpretable model around a specific prediction. It works by perturbing the input and fitting a simple model, like a linear model, to explain the behaviour of the black box model in the vicinity of the prediction of interest. LIME  approximates model behavior locally around ($x$) by minimizing: $\underset{g \in G}{\operatorname{argmin}} L(\mathcal{M}, g, \pi_x) + \Omega(g)$ where $g$ is an interpretable model in the neighborhood of ($x$).

\subsubsection{Integrated Gradients}
Integrated Gradients (IG)~\cite{sundararajan2017axiomatic} attributes the prediction of a deep network to its inputs by integrating the gradients along a straight-line path from a baseline input to the actual input. This method satisfies desirable axioms like completeness and sensitivity, providing a theoretically sound approach to feature attribution. IG attributes feature importance by integrating model gradients from a baseline   $  {\bf g} ({\bf \mathcal{M}}, x) = \text{IG}(x)=(x - \bar{x}) \times \int_{\alpha=0}^1 \frac{{\partial {\bf \mathcal{M}}(\bar{x} + \alpha \cdot (x - \bar{x}))}}{{\partial x}} \, d\alpha $ where \( \bar{x} \) is the baseline input.

\subsubsection{DeepLIFT}
DeepLIFT \cite{shrikumar2017learning} assigns each input ($x$) a value  \(C_{\Delta x_i \Delta y}\) representing its deviation from a reference value, satisfying: $\sum_{i=1}^{n} C_{\Delta x_i \Delta o} = \Delta o $
where  \(o = \mathcal{M}(x)\) and \(\Delta o\) is the difference between model output and reference value.

\subsection{Evaluation of Explainable AI methods}\label{sec:evaluate_xai}
The evaluation of Explainable AI methods is crucial to ensure that the explanations they provided are  transparent,  also accurate and reliable. We employ four key metrics to assess the quality of our explanations for LSTM Model based NIDS alerts: Reliability,  Faithfulness, Robustness and Complexity. These metrics provide a comprehensive evaluation framework that addresses different aspects of explanation quality. XAI evaluation is categorized into three groups~\cite{coroama2022evaluation}: user-focused evaluation, application-focused evaluation, and functionality-focused evaluation. The first two categories are part of human-centered evaluation and are broken down into subjective and objective measures.

\subsubsection{Reliability} An explanation should be centered around the region of interest, the ground truth $\mathbf{GT}$. \(\mathbf{g}(\mathcal{M}, \mathbf{x}) = \mathbf{GT}\). 'Major' parts of an explanation should lie inside the ground truth mask GT(x) for both Relevance Mass Accuracy and Relevance Rank Accuracy metrics used in this work, and the Ground truth mask ([0,1]) was determined by the features SOC Analysts identified (see Table.~\ref{tab:soc_features}).Truth-based measures relevance rank accuracy and relevance mask accuracy are derived from~\cite{arras2022clevr}.

\begin{enumerate}[label=(\alph*), leftmargin=*]
    \item Relevance Rank Accuracy (RRA)~\cite{arras2022clevr}:  Relevance rank accuracy measures how much of the high-intensity relevance lies within the ground truth. We sort the top \(K\) values of \(\mathbf{g}(\mathcal{M}, \mathbf{x})\) in decreasing order \(\mathbf{X}_{\text{top}K} = \{x_1, ..., x_K \mid \mathbf{g}(\mathcal{M}, \mathbf{x})_{x_1} > ... > \mathbf{g}(\mathcal{M}, \mathbf{x})_{x_K}\}\).
    \[
    RRA = \frac{|\mathbf{X}_{\text{top}_k} \cap \mathbf{GT}(\mathbf{x})|}{|\mathbf{GT}(\mathbf{x})|}
    \]
    Here $\text{top}_k$ are features Identified by SOC Analyst. 
    \item Relevance Mass Accuracy (RMA)~\cite{arras2022clevr}: The relevance mass accuracy is calculated as the sum of the explanation values within the ground truth mask divided by the sum of all values. 

        \[
            \text{RMA} = \frac{\sum_{i} \mathbf{g}(\mathcal{M}, \mathbf{x})_i \cdot \mathbf{GT}(x_i)}{\sum_{i} \mathbf{g}(\mathcal{M}, \mathbf{x})_i}
        \]
    
\end{enumerate}

\subsubsection{Faithfulness} The explanation algorithm $\mathbf{g}$ should replicate the model's behavior. \(\mathbf{g}(\mathcal{M}, \mathbf{x}) \approx \mathcal{M}(\mathbf{x})\). Faithfulness quantifies the consistency between the prediction model  $\mathcal{M}$ and explanation $g$. For evaluating the Faithfulness of explanations, the Faithfulness correlation~\cite{bhatt2020evaluating} and Monotonocity~\cite{luss2019generating} metrics were used. 

\begin{enumerate}[label=(\alph*), leftmargin=*]
    \item High Faithfulness Correlation:
        Faithfulness measures how well the explanation function ${\bf g}$ aligns feature importance scores with the black-box model  ${\mathcal{M}}$
    \small
    \begin{equation}
        \mu_F(\mathcal{M},g;x) = \underset{{\mathcal B} \in \binom{|d|}{|{\mathcal B}|}}{\text{corr}} \Biggl( \sum_{i \in {\mathcal B}} g(\mathcal{M},x)_i, \mathcal{M}(x) - \mathcal{M}(x_{\mathcal B}) \Biggr)
    \end{equation}
    \normalsize
        where $x_{\mathcal B} = x_i | i \in {\mathcal B}\}$
        High Faithfulness correlation metric iteratively substitutes a random subset of given attributions with a baseline value $\mathcal B$. Then, it measures the correlation between the sum of these attributions and the difference in the model's output. 
    
    \item Monotonicity: 
        Let $x, x' \in \mathcal{R}^d$ be two input points such that $x_i \leq x'_i$ for all $i \in {1,2,\dots,d}$. $\mathcal{M}$ and $g$ are said to be monotonic if the following condition holds: For any subset $S \subseteq {1,2,\dots,d}$, the sum of the attributions of the features in $S$ should be nonnegative when moving from $x$ to $x'$, that is, $\sum_{i \in S} g(\mathcal{M},x)i \leq \sum_{i \in S} g(\mathcal{M},x')_i$ implies $$\mathcal{M}(x) - \mathcal{M}(x_{[x_s=\bar{x}_s]}) \leq \mathcal{M}(x') - \mathcal{M}(x'{[x'_s=\bar{x}_s]})$$
\end{enumerate}

\subsubsection{Robustness} Robustness refers to similar inputs should result in similar explanations. \(\mathbf{g}(\mathcal{M}, \mathbf{x}) \approx \mathbf{g}(\mathcal{M}, \mathbf{x} + \epsilon) \text{ for small } \epsilon\). 

\begin{enumerate}[label=(\alph*),  leftmargin=*]
    \item Max Sensitivity: Max sensitivity~~\cite{bhatt2020evaluating}: is used to ensure that nearby inputs with similar model output have similar explanations, it is desirable for the explanation function $g$ to have a low sensitivity in the region surrounding the point of interest $x$, assuming the differentiability of the predictor function $\mathcal{M}$. Maximum sensitivity of an explanation function $g$ at a point of interest $x$ in its neighbourhood is defined as follows: Consider a neighbourhood $N_r$ of points within a radius $r$ of $x$, denoted by $N_r={z \in D_x |p(x,z) \leq r, \mathcal{M}(x)=\mathcal{M}(x)(z)}$, where $D$ is the distance metric, and $p$ is the proximity function. Given a predictor $\mathcal{M}(x)$, a distance metric $D$, a proximity function $p$, a radius $r$, and a point $x$, we define the maximum sensitivity of $g$ at $x$ as follows:
          \small
        \begin{equation}
            \mu_M(\mathcal{M}(x),g,r;x) = \underset{z \in N_r}{\operatorname{max}} D\left( g(\mathcal{M}(x),x), g(\mathcal{M}(x),z) \right)
        \end{equation}
        \normalsize
\end{enumerate}

\subsubsection{Complexity} Explanations using a smaller number of features are preferred. It is assumed that explanations using a large number of features are difficult for the user to understand.\(\min \left\|\mathbf{g}(\mathcal{M}, \mathbf{x})\right\|_0\).  
\begin{enumerate}[label=(\alph*)]
    \item Low Complexity: Low complexity~\cite{bhatt2020evaluating} metric computes the entropy of each feature's fractional contribution to the total attribution magnitude individually. 
    \begin{equation}
        \mu_C(\mathcal{M},g;x) = - \sum_{i = 1}^d P_g(i) \log P_g(i)
        \end{equation}
        where 
        \begin{equation}
        P_g(i) = \frac{|g(\mathcal{M},x)i|}{\sum{j\in |d|} |g(\mathcal{M},x)_j|}; P_g = {P_g(1),.... P_g(d)}
    \end{equation}
\end{enumerate}

The experiments were carried out on a computer running Pop!\_OS 22.04 LTS x86\_64 operating system with the following hardware configuration: 32 GB of DDR4-2666R ECC RAM, AMD Ryzen 5 5600G with Radeon Graphics (12) @ 3.900GHz processor. The scripts were developed using the Python 3.9 programming language and Pytorch library. For the implementation of the Integrated Gradients and DeepLIFT explainers, Captum library was used.

\section{Results \& Discussions}\label{sec:results_discussion}
In this section, we present the results of our research, including an analysis of the LSTM model's performance and explanations of LSTM model using Explainable AI methods and the quality of evaluation for these explanations based on four criteria: faithfulness, complexity, reliability, and robustness.  

\begin{figure}[t]
    \centering
    \begin{subfigure}[t]{0.45\linewidth}
        \centering
        \includegraphics[width=\linewidth]{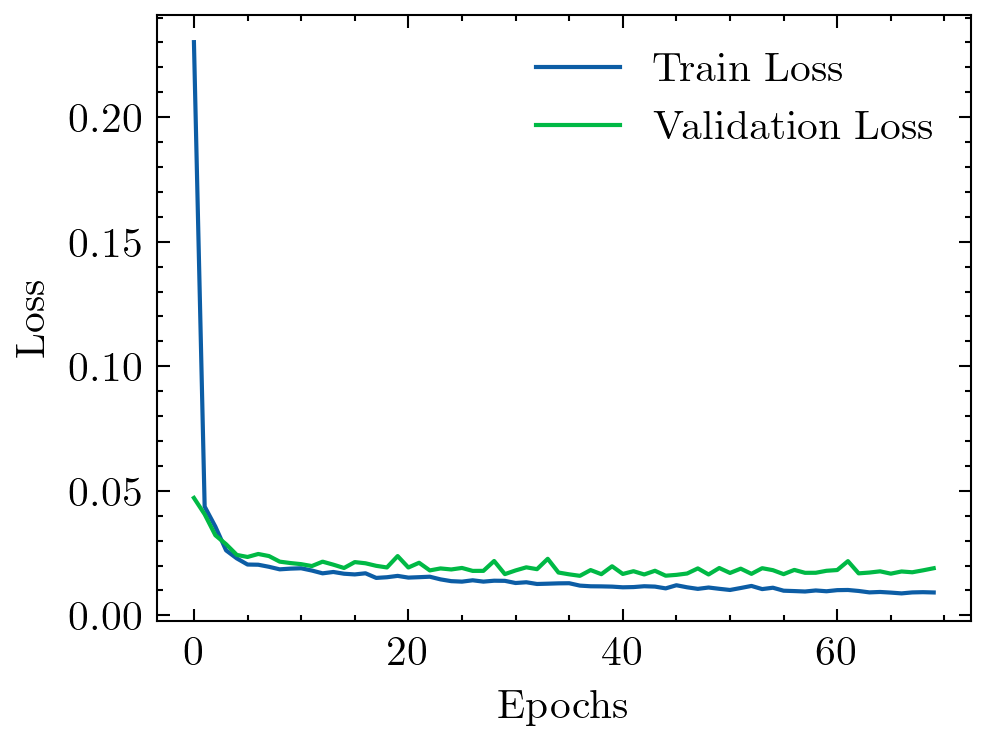}
        \caption{Loss}
        \label{fig:loss}
    \end{subfigure}
    \vspace{1em}
    \begin{subfigure}[t]{0.45\linewidth}
        \centering
        \includegraphics[width=\linewidth]{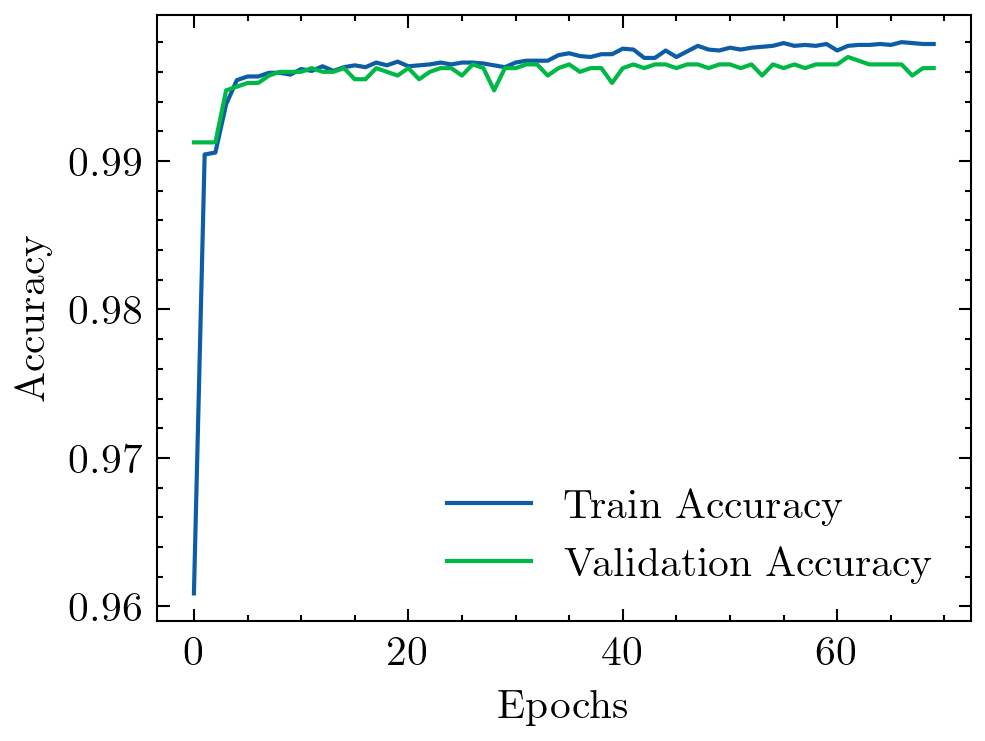}
        \caption{Accuracy.}
        \label{fig:accuracy}
    \end{subfigure}
    \caption{Loss and Accuracy from Best LSTM Performance Model}
    \label{fig:loss_and_accuracy}
\end{figure}

\begin{figure}[t]
    \centering
    \begin{subfigure}[t]{0.45\linewidth}
        \centering
        \includegraphics[width=\linewidth]{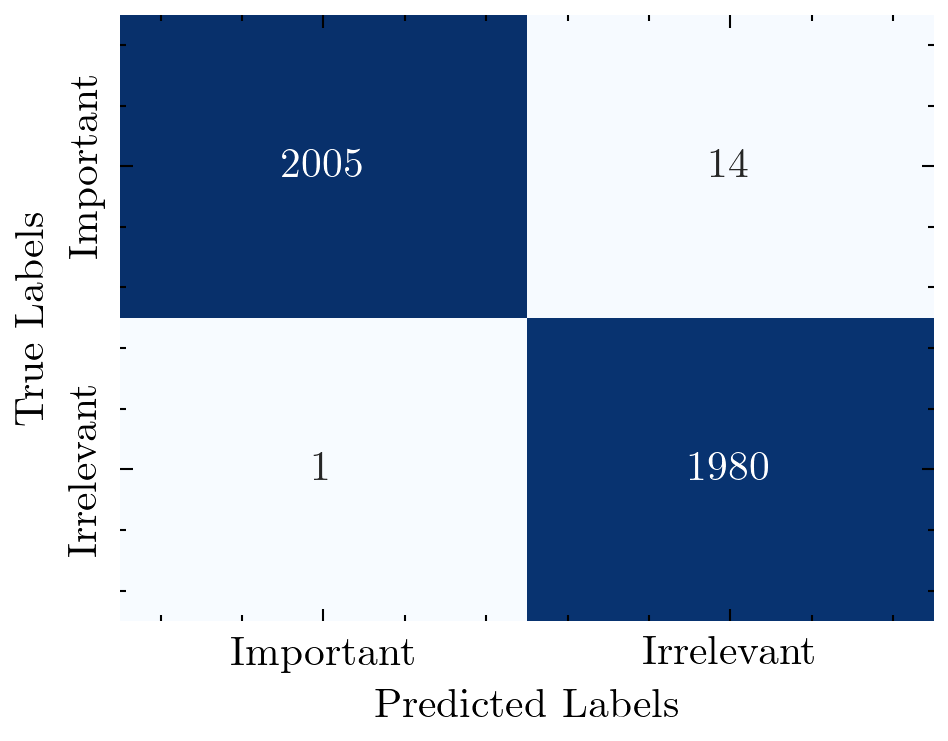}
        \caption{Confusion Matrix}
        \label{fig:confusion_matrix}
    \end{subfigure}
    \vspace{1em}
    \begin{subfigure}[t]{0.45\linewidth}
        \centering
        \includegraphics[width=\linewidth]{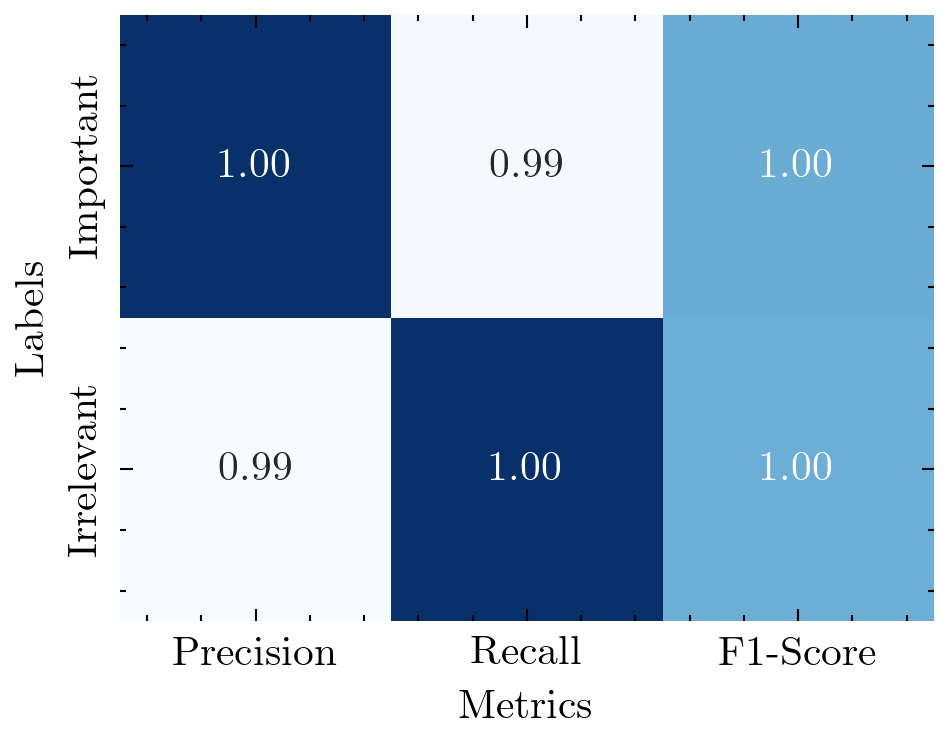}
        \caption{Classification Report}
        \label{fig:classification_report}
    \end{subfigure}
    \caption{Confusion Matrix and Classification Report}
    \label{fig:cm_clf_report}
\end{figure}

Figure.~\ref{fig:confusion_matrix} shows the confusion matrix, indicating the model's strong classification performance for test data of 4000. It correctly classified 2005 important alerts and 1980 irrelevant alerts, with only 14 misclassifications of irrelevant alerts as necessary, demonstrating high accuracy and a low false positive rate. Figure.~\ref{fig:loss} shows the training and validation loss over 70 epochs obtained through random search parameter tuning. Initially, both decrease rapidly before stabilizing, indicating convergence without overfitting. The close alignment of the training and validation loss curves represents good generalization to unseen data. Figure.~\ref{fig:accuracy} shows the training and validation accuracy, which quickly stabilizes above 99.5\%, indicating strong model performance. In Figure.~\ref{fig:classification_report}, from the classification report,  the model achieves near-perfect precision, recall, and F1-score scores for both classes.

\begin{figure}[htbp]
    \centering
    \begin{subfigure}[b]{0.45\textwidth}
        \centering
        \includegraphics[width=\textwidth]{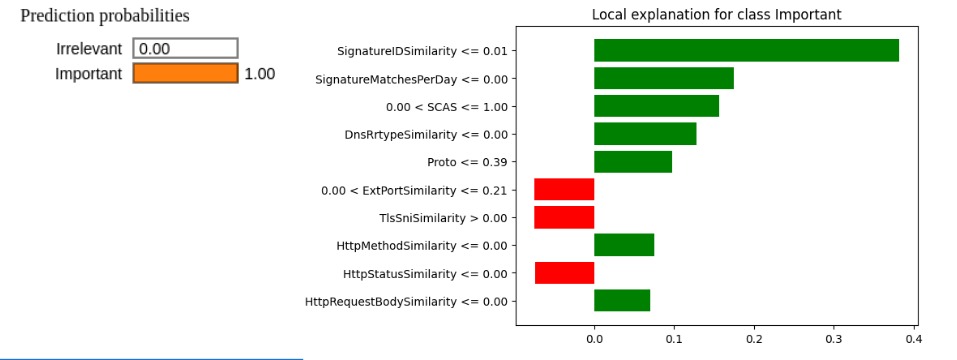}
        \caption{LIME explanations for important NIDS alerts using an LSTM model}
        \label{fig:lime_exp}
    \end{subfigure}
    \hfill
    \begin{subfigure}[b]{0.45\textwidth}
        \centering
        \includegraphics[width=\textwidth]{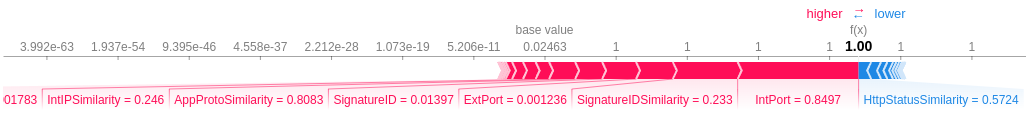}
        \caption{SHAP explanations for an important NIDS alert data point using an LSTM model}
        \label{fig:shap_exp}
    \end{subfigure}
    \hfill
    \begin{subfigure}[b]{0.45\textwidth}
        \centering
        \includegraphics[width=\textwidth]{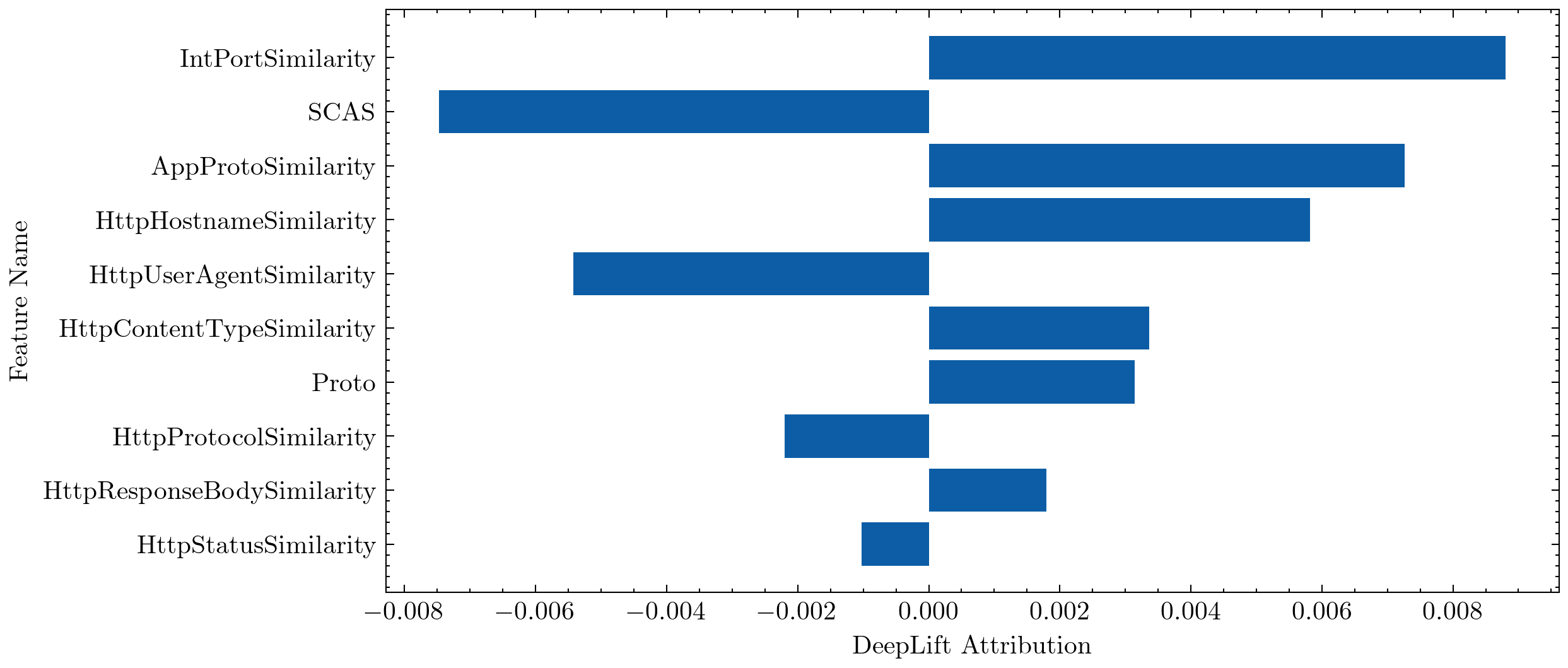}
        \caption{DeepLIFT feature importance for an important NIDS alert data point using an LSTM model}
        \label{fig:deept_exp}
    \end{subfigure}
    \hfill
    \begin{subfigure}[b]{0.45\textwidth}
        \centering
        \includegraphics[width=\textwidth]{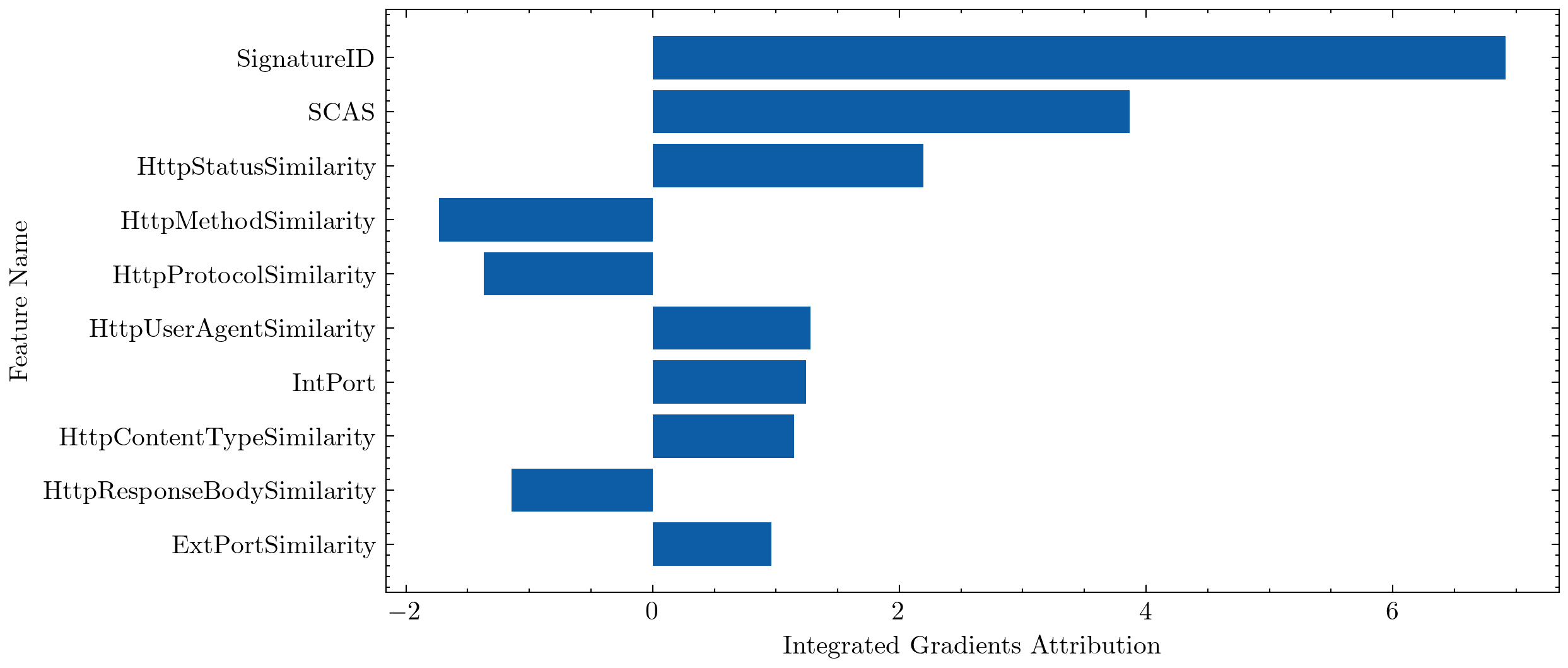}
        \caption{Integrated Gradients feature importance for an important NIDS alert data point using an LSTM model}
        \label{fig:ig_exp}
    \end{subfigure}
    \caption{Explanations for an important NIDS alert data point using an LSTM model}
    \label{fig:explanations_graphs}
\end{figure}

In this paper, we utilized 4 different explainable AI methods (LIME, SHAP, IG, and DeepLift) to explain the predictions of our LSTM model on the test data. LIME analyzes how the model assigns probabilities to categories by comparing these probabilities with the actual category of the data point.  SHAP method provides single-data-point explanations for models, giving insights. In explanations, a particular data point is selected to demonstrate how each feature influences the model's prediction. 

Fig.~\ref{fig:lime_exp} shows an local explanation from LIME method for a NIDS alert labeled as "Important.".  Left side presents prediction probabilities with a 100\% probability for the "Important" class. On the right side it illustrates the impact of features. For instance, when the feature `SignatureIDSimilarity' is less than or equal to 0.01, it positively affects the "Important" classification of NIDS alert. Additionally, `SignatureMatchesPerDay' and `SCAS' being less than or equal to 1.00 also contribute positively. Conversely, `ExtPortSimilarity' and `TlsSniSimilarity' have impacts, suggesting that some NIDS alerts may not be relevant. SHAP employs Shapley values to showcase how features influence model predictions in Fig.~\ref{fig:shap_exp}  of force plot, red bar signifies the positive impact while blue bar indicates the negative impact on the model output. Each bar demonstrates whether the features bring the predicted value closer to or farther from the base value of 0.02463. The plot's base value is the average of all prediction values. Each strip in the plot displays the impact of the features on moving the predicted value closer to or farther from the base value. Final prediction is deemed an "important class label", with a value of 1.00 for this NIDS alert. Features, like 'IntPort' (Internal Port) 'SignatureIDSimilarity'. ExtPort' (External Port) along with 'SignatureID' play a role in indicating the importance of NIDS alert. However, the feature 'HttpStatusSimilarity' might suggest that this alert could be a less critical feature to its impact.

DeepLift is a technique used to attribute the output of LSTM model to its input features by comparing neuron activation to a reference activation and assigning contribution scores based on the variance. Fig.~\ref{fig:deept_exp} illustrates the significance of features using the DeepLift explainer for the 10 features of a NIDS alert data point labeled as "important."  The negative attribution of 'SCAS' suggests its influence on classifying as "Important" in NIDS alerts. Additionally 'HttpMethodSimilarity' and 'IntIP' show negative attributions while HttpContentTypeSimilarity has a slight positive impact countering the "Important" classification. IG attribute a LSTM model's prediction its input features by integrating gradients of the model's output with respect to the input along from a baseline to the input. This explanation technique works best for models that use linear activation functions. Fig.~\ref{fig:ig_exp} showcases feature importance using IG explainer for a data point in the "Important" NIDS alert class label among the 10 features. Features such, as 'SignatureID' 'SCAS,' and 'HttpStatusSimilarity' display attributions.

\begin{table*}[ht]
\centering
\caption{Evaluation Results of Explainable AI Methods: Mean ($\mu$) and Standard Deviation ($\sigma$) Values.}
\resizebox{\textwidth}{!}{%
\begin{tabular}{@{}cccccccccc@{}}
    \toprule
    Explanation Criterion & \multicolumn{2}{c}{Faithfulness}                                               &  & Robustness               &  & Complexity               &  & \multicolumn{2}{c}{Reliability}                                                                                                         \\ \midrule
    Explainer/Metric      & \begin{tabular}[c]{@{}c@{}}High\\  Faithfulnes\end{tabular} & Monotonicity     &  & Max Sensitvity           &  & Low Complexity           &  & \begin{tabular}[c]{@{}c@{}}Relevance Mass\\  Accuracy\end{tabular} & \begin{tabular}[c]{@{}c@{}}Relevancy Rank\\  Accuracy\end{tabular} \\
                          & $\mu \pm \sigma $                                                      &   $\mu$               &  & $\mu \pm \sigma $                   &  & $\mu \pm \sigma $                     &  & $\mu \pm \sigma $                                                               &$\mu \pm \sigma $                                                               \\ \cmidrule(l){2-10} 
    Lime                  & 0.4209 ± 0.1835                                             & 59.55\%          &  & 0.3617 ± 0.1152          &  & 3.0318 ± 0.0703          &  & 0.6234 ± 9.7008                                                    & 0.5250 ± 0.1041                                                    \\
    Shap                  & 0.3959 ± 0.2928                                             & 64.45\%          &  & 0.0245 ± 0.0862          &  & 2.4677 ± 0.2074          &  & 0.6527 ± 3.8334                                                    & 0.4743 ± 0.1418                                                    \\
    IG                    & 0.1761 ± 0.3815                                             & 73.70\%          &  & 0.1774 ± 0.2505          &  & \textbf{2.1745 ± 0.4134} &  & 0.5939 ± 0.6840                                                    & 0.3410 ± 0.1545                                                    \\
    Deep Lift             & \textbf{0.7559 ± 0.2681}                                    & \textbf{78.35\%} &  & \textbf{0.0008 ± 0.0004} &  & 2.2635 ± 0.3299          &  & \textbf{0.7812 ± 25.2805}                                          & \textbf{0.6754 ± 0.0897}                                           \\ \bottomrule
    \end{tabular}
}
\label{tab:xai_eval_results}
\end{table*}

\begin{figure*}
\centering
    \begin{subfigure}[t]{0.24\textwidth}
        \centering
        \includegraphics[width=\textwidth]{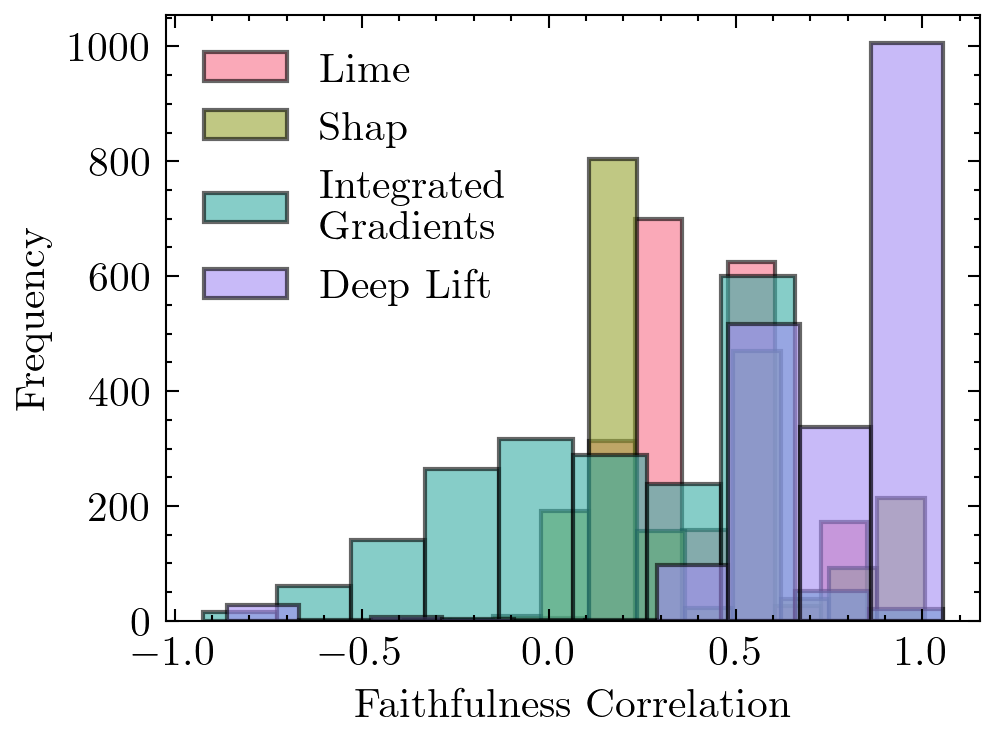}
        \caption{High Faithfulness} 
        \label{fig:faith_corr}
    \end{subfigure}\hfill
    \begin{subfigure}[t]{0.24\textwidth}
        \centering
        \includegraphics[width=\textwidth]{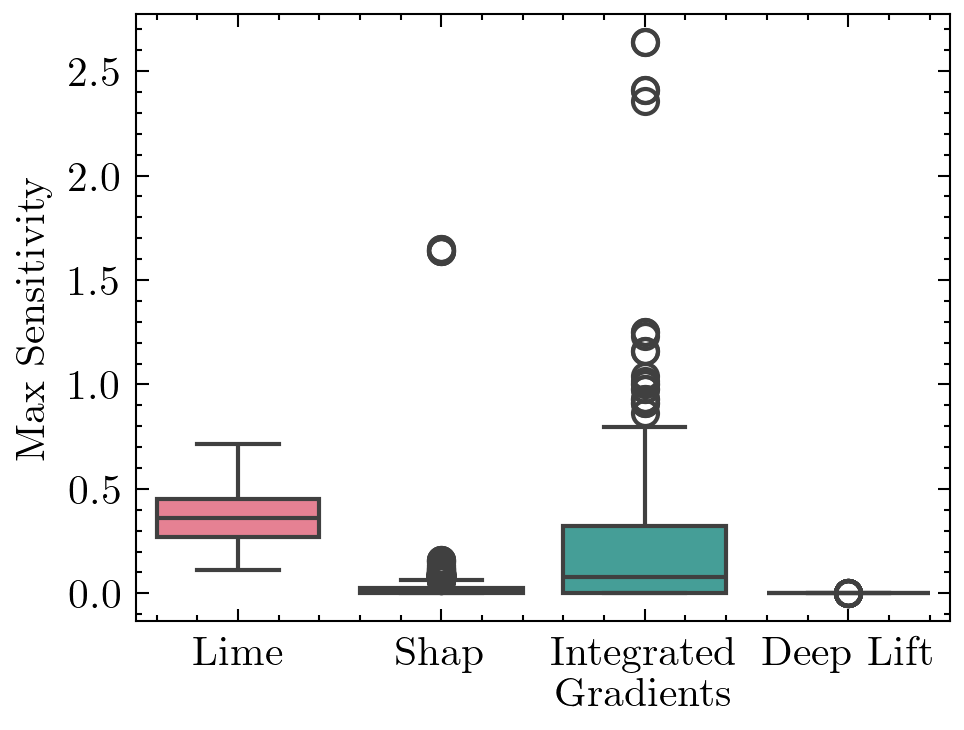}
        \caption{Max Sensitivity} 
        \label{fig:max_sens}
    \end{subfigure}\hfill
    \begin{subfigure}[t]{0.24\textwidth}
        \centering
        \includegraphics[width=\textwidth]{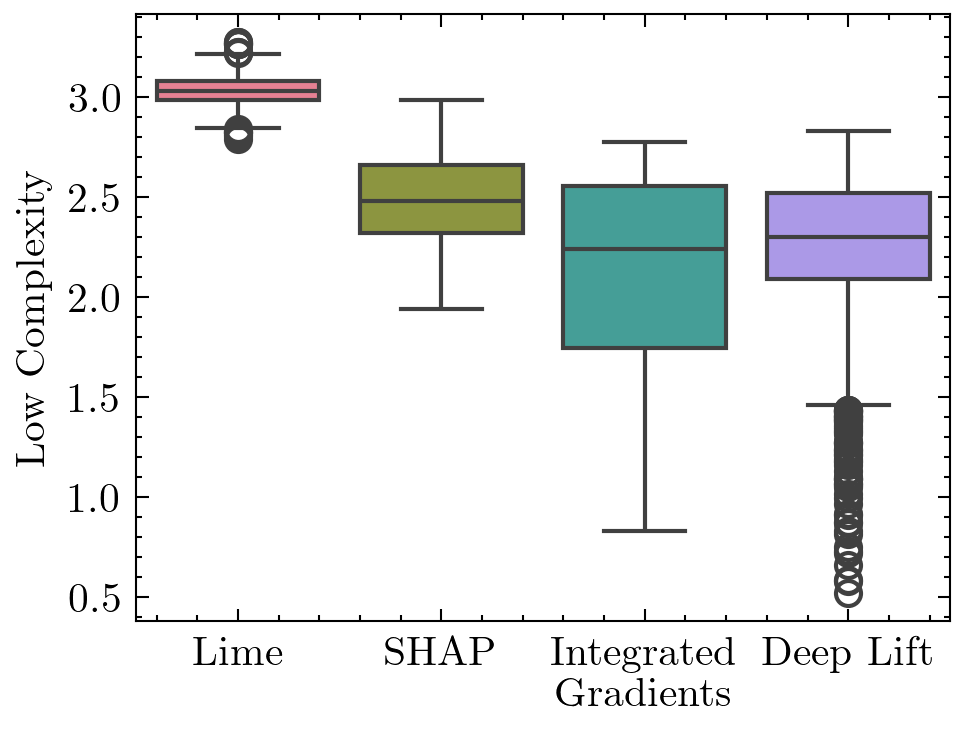}
        \caption{Low Complexity} 
        \label{fig:low_complexity}
    \end{subfigure}\hfill
    \begin{subfigure}[t]{0.24\textwidth}
        \centering
        \includegraphics[width=\textwidth]{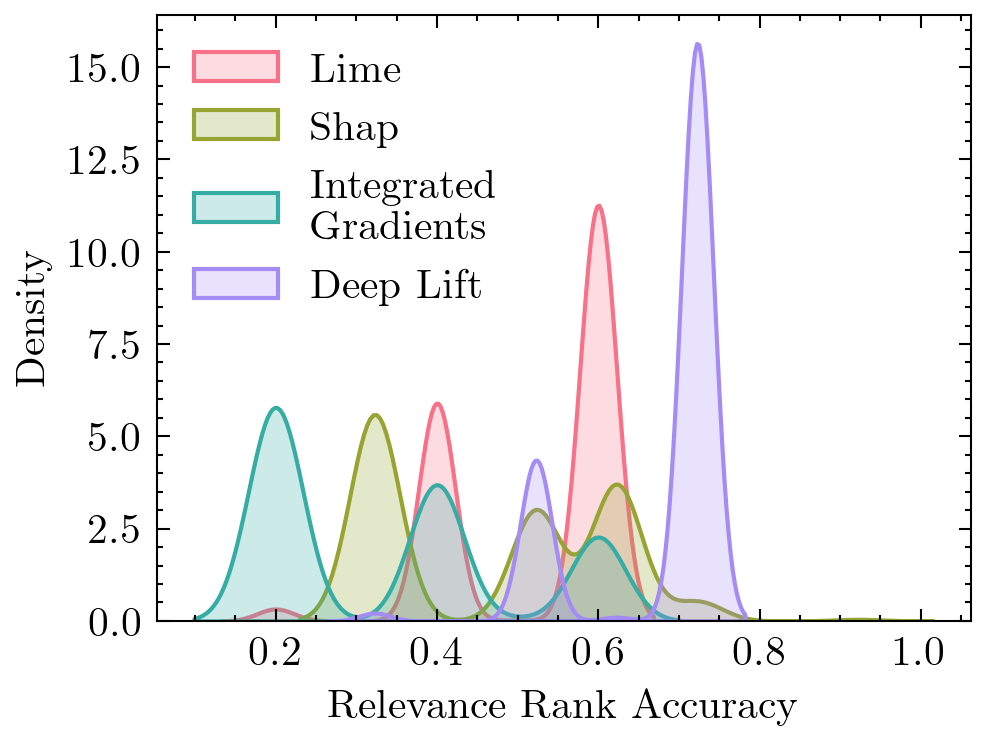}
        \caption{Relevancy Rank Accuracy} 
        \label{fig:rra}
    \end{subfigure}
    \caption{Quality of Explainable AI evaluation metrics distribution}
    \label{fig:xai_metrics_eval_graph}
\end{figure*}

\begin{table*}[ht]
\centering
\footnotesize
\caption{Statistical Comparison of Explainers Across Multiple Metrics ($p$-values)}
\begin{tabular}{@{}ccccc@{}}
\toprule
Metric                           & Explainer & Shap         & IG            & Deep Lift     \\ \midrule
\multirow{3}{*}{Faithfulness}    & LIME      & L (3.34e-41) & L (1.03e-134) & D (5.61e-185) \\
                                 & SHAP      & -            & S (6.22e-91)  & D (1.03e-169) \\
                                 & IG        &              & -             & D (1.30e-230) \\ \midrule
\multirow{3}{*}{Max Sensitivity} & LIME      & S (0.00e+00) & I (1.64e-221) & D (0.00e+00)  \\
                                 & SHAP      & -            & S (1.38e-185) & D (3.54e-126) \\
                                 & IG        &              & -             & D (3.29e-126) \\ \midrule
\multirow{3}{*}{Low Complexity}  & LIME      & S (0.00e+00) & I (0.00e+00)  & D (0.00e+00)  \\
                                 & SHAP      & -            & I (5.45e-146) & D (1.26e-88)  \\
                                 & IG        &              & -             & I (1.07e-42)  \\ \midrule
\multirow{3}{*}{RMA}             & LIME      & S (5.12e-25) & L (1.97e-80)  & D (6.22e-83)  \\
                                 & SHAP      & -            & S (4.67e-155) & D (6.47e-91)  \\
                                 & IG        &              & -             & D (2.82e-54)  \\ \midrule
\multirow{3}{*}{RRA}             & LIME      & L (7.61e-39) & L (3.07e-210) & D (0.00e+00)  \\
                                 & SHAP      & -            & S (5.52e-155) & D (3.97e-253) \\
                                 & IG        &              & -             & D (0.00e+00)  \\ \bottomrule
\end{tabular}

\begin{tablenotes}
\centering
\footnotesize
\item[1] D (Deep Lift), L (LIME), S (SHAP), and I (Integrated Gradients) \\ indicate the better performing explainer in each pairwise comparison.
\item[$\bullet$] $p > 0.05$ — No significant evidence against $H_0$; $H_0$ is not rejected
\item[$\bullet$] $0.01 < p \leq 0.05$ — Significant evidence against $H_0$; $H_1$ is accepted at 95\% confidence level
\item[$\bullet$] $0.001 < p \leq 0.01$ — Strong evidence against $H_0$; $H_1$ is accepted at 99\% confidence level
\item[$\bullet$] $p \leq 0.001$ — Very strong evidence against $H_0$; $H_1$ is accepted at 99.9\% confidence level. 
\end{tablenotes}
\label{tab:stat_analysis}
\end{table*}

Our analysis comparing the features identified by the TalTech SOC analyst closely aligned with those derived by explainers used in our LSTM model to classify "important" NIDS alerts. The 5 features recognized by SOC experts in Table~\ref{tab:soc_features} proved significant across explainers, although their order of feature importance varied. For instance, 'SignatureIDSimilarity' and 'SignatureID', highlighted by SOC analysts, impacted the SHAP explainer for NIDS alerts. The presence of "SCAS" was notable in LIME, IG, and DeepLift, confirming its significance. The importance of 'SignatureMatchesPerDay' varied among explainers within LIME. Notably upon reviewing the 10 features highlighted by each explainer, we noticed an overlap with the features identified by SOC analysts particularly emphasizing 'SignatureID', 'SignatureIDSimilarity', 'SCAS' and 'SignatureMatchesPerDay'. We assessed the quality explanation of XAI methods, for LSTM model based alerts using metrics based on four criteria: faithfulness, robustness, complexity and reliability.


We evaluated the quality of explanations obtained by XAI methods for  Long Short-Term Memory (LSTM) network-based NIDS alert classification across 2000 data points using metrics based on four criteria: Faithfulness, robustness, complexity, and reliability.  Table~\ref{fig:xai_metrics_eval_graph} shows the results of the quality of explanation for XAI methods. LSTM model prediction probabilities were computed using the Softmax activation function. To evaluate the Faithfulness of explanations, we employed high faithfulness correlations and monotonicity.  High Faithfulness of XAI methods was evaluated by studying the correlation between attribute importance assigned by the XAI method and their impact on the model's probabilities.  A high faithfulness correlation value suggests that the explanations effectively capture the model's behaviour and can be regarded as faithful. Table.~\ref{tab:xai_eval_results} shows the evaluation results of xai methods.  Mean ($\mu$) and standard deviation ($\sigma$) values were calculated for the test data of XAI computed metrics for 2000 test data points.   Deep Lift achieved the highest Faithfulness mean and standard deviation correlation values of 0.7559 ± 0.2681 for test data points.  We also analyzed the monotonicity of the explanation to understand how individual features affect model probability by adding each attribute to enhance its importance and observing its influence on the model's probability. By assessing the monotonicity of the explainer, we can measure how the explanations change monotonically with respect to the input features. Deep LIFT achieved high monotonicity with 78\% ($\mu$). 

To measure complexity, we calculate the entropy of feature attribution in the explanations. Complexity measures the conciseness of explanations derived by the explainer. Among xai methods assessed by low complexity metric,  Integrated Gradients (IG) achieved lower complexity ( 2.174 ± 0.413) closely followed by DeepLift  ( 2.264 ±
0.330.) 

The sensitivity metric assesses the consistency of the explainers' output, ensuring that similar inputs in the feature space of model outputs have similar explanations when sensitivity is low.  For this metric, we used the Euclidean distance with a radius value of 0.1 to find the nearest neighbour points related to the prediction label of an explanation which helps to identify data points in the feature space with similar explanations for the predicted label. Deep LIFT achieved Lower sensitivity with max sensitivity metric (0.0008 ± 0.0004). 

Two metrics, Relevance Mass Accuracy and Relevance Rank Accuracy, were used to evaluate the reliability of explanations. These metrics validated the explanations by comparing them to a ground truth mask based on features identified through collaboration with an SoC analyst. For both Relevance Mass Accuracy (0.781 ± 25.281) and Relevancy Rank Accuracy (0.6754 ± 0.089) metrics, Deep lift explanations were reliable.  Figure.~\ref{fig:xai_metrics_eval_graph} illustrates the distribution of XAI metric results for 2000 data points, highlighting that DeepLIFT's explanations demonstrate high faithfulness, lower sensitivity, lower complexity, and more relevance rank accuracy. Faithfulness correlation values for DeepLIFT indicate a strong skew towards higher levels, showing a high degree of consistency through monotonicity. Moreover, the entropy values of feature importance scores for IG and DeepLIFT are more evenly spread towards the lower end than other explainers. The sensitivity values for the DeepLIFT explainer are also more evenly spread to lower values in maximum sensitivity metrics. Additionally, using Relevance Rank Accuracy, DeepLIFT consistently achieves a high relevance rank accuracy with less variation, centred around 0.8.  

Following established practices in the statistical analysis of XAI methods evaluation~\cite{jesus2021can}, we employed the Wilcoxon signed-ranks test~\cite{woolson2005wilcoxon} to evaluate the statistical significance of differences~\cite{demvsar2006statistical} in XAI metric scores between pairs of explainers (i.e., $explainer_A$, $explainer_B$) for NIDS alert classification. The null hypothesis ($H_0$) is that the explainable AI metric scores of the explainers are equivalent, i.e., there is no significant difference between the explainers (XAI Metric Score($explainer_A$) = XAI Metric Score($explainer_B$)). The alternative hypothesis ($H_1$) is that they are not equivalent (XAI Metric Score($explainer_A$) $\neq$ XAI Metric Score($explainer_B$)), indicating a significant difference in their explainer metric scores. XAI metrics used in this study are High Faithfulness, Max Sensitivity, Low Complexity, Relevance Mass Accuracy, and Relevancy Rank Accuracy. This test was conducted separately for each metric to assess the performance differences among the explainers comprehensively.

The statistical analysis in Table ~\ref{tab:stat_analysis} shows significant differences among the explainers for all metrics, with p-values consistently below 0.05, demonstrating strong evidence against the null hypothesis. DeepLift explainer is better regarding faithfulness, max sensitivity, RMA, and RRA when compared pairwise ($p < 0.001$ for all comparisons) with other explainers. The relative performance of SHAP, LIME, and IG varies across metrics can be seen Table~\ref{tab:stat_analysis}.

\begin{figure}[ht]
    \centering
    \setkeys{Gin}{width=\linewidth}
    \includegraphics{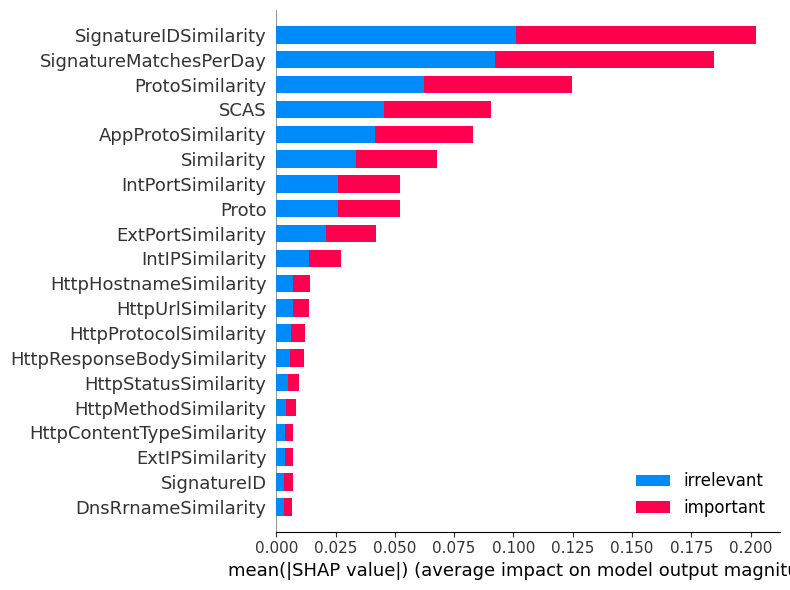}
    \caption{SHAP global explanation for LSTM model}
    \label{fig:global}
\end{figure}

We have also provided a global explanation using SHAP values for all the testing data of the LSTM model. A higher value positively impacts the prediction, while a lower value contributes negatively. Figure.~\ref{fig:global} shows the global explanation of the LSTM model.  The graph illustrates the average impact of each feature on the model's output magnitude for the class labels,  "irrelevant" and "important" classifications. SignatureIDSimilarity, SignatureMatchesPerDay, ProtoSimilarity and SCAS are most impact ful features for important nids alerts. Notably, these top features align with those identified by human expert SOC analysts.     Lower-ranked features such as HTTP-related similarities (e.g., HttpHostnameSimilarity, HttpUrlSimilarity) and IP-related features (e.g., ExtIPSimilarity) have comparatively less impact on the model's decisions.

\section{Conclusions and Future work}\label{sec:conclusions}
This research presents explainable artificial intelligence (XAI) based Network Intrusion Detection Systems (NIDS) alert classification utilizing a Long Short-Term Memory (LSTM) model. We have showcased how enhancing the explainability and trustworthiness of AI-powered cybersecurity systems can be achieved by clarifying the output predictions of these LSTM models through four XAI techniques: LIME, SHAP, Integrated Gradients, and DeepLIFT. Our thorough assessment of the XAI framework, considering the aspects of faithfulness, complexity, robustness, and reliability, has evaluated how well these XAI methods explain NIDS alerts. The superior performance of DeepLIFT across these evaluation metrics underscores its potential as a preferred method for interpreting NIDS alert classifications. Notably, the substantial alignment between explanations generated by XAI techniques and features identified by SOC analysts validates their effectiveness in capturing domain expertise. This research makes a contribution by bridging the gap between the high accuracy of opaque machine learning models and the necessity for transparent decision-making in cybersecurity operations. By proposing a framework to explain black box model decisions and assess XAI in NIDS applications, we provided comprehensive benchmarking results, including evaluation metrics for developing transparent and interpretable AI systems in crucial security domains.

\bibliographystyle{unsrt}  


\end{document}